\documentclass[showpacs]{revtex4}
\usepackage{graphicx}
\usepackage{amsmath}

\usepackage{amsmath}
\usepackage{amsfonts}

\usepackage{epsfig} 

\usepackage{epstopdf}
\newcommand{\Ket}[1]{\left|#1  \right>}

\newcommand{\Braket}[1]{\left<#1  \right>}

\newcommand{\trace}{{\mathrm{Tr}}}

\def\begeq{\begin{equation}}
\def\endeq{\end{equation}}

\def\begeqar{\begin{eqnarray}}
\def\endeqar{\end{eqnarray}}

\usepackage{color}

\begin{document}

\title{Entanglement in quantum impurity problems is non perturbative}

\author{H. Saleur}
\affiliation{Institut de Physique Th\'eorique, CEA, IPhT and CNRS, URA2306, Gif Sur Yvette, F-91191}
\affiliation{Department of Physics, University of Southern California, Los Angeles, CA 90089-0484}
\author{P. Schmitteckert}
\affiliation{Institute for Nanotechnology, 
             Karlsruhe Institute of Technology,
             76344 Eggenstein-Leopoldshafen, Germany}
\author{R. Vasseur}
\affiliation{Institut de Physique Th\'eorique, CEA Saclay,
91191 Gif Sur Yvette, France}
\affiliation{LPTENS, 24 rue Lhomond, 75231 Paris, France}


\date{\today}

\begin{abstract}

We study  the  entanglement entropy of a region of length $2L$ with the remainder of an infinite one dimensional gapless quantum system in the case where the region is centered on a quantum impurity. The coupling to this impurity is not scale invariant, and the physics involves a crossover between weak and strong coupling regimes. While the impurity contribution to the entanglement has been computed numerically in the past, little is known analytically about it, since in particular the methods of conformal invariance cannot be applied because of the presence of a crossover length. 

We show in this paper that the small coupling  expansion of the entanglement entropy in this problem is quite generally  plagued by strong infrared divergences, implying a  non-perturbative dependence on the coupling. The  large coupling expansion turns out to be better behaved, thanks to powerful results from  the boundary CFT formulation and, in some cases,  the underlying integrability of the problem.  However, it is clear that this expansion does not capture well the crossover physics. 

In the integrable case -- which includes problems such as an XXZ chain with a modified link, the interacting resonant level model or the anisotropic Kondo model -- a non perturbative approach is in principle possible using form-factors.  We adapt in this paper the ideas of~\cite{Doyon1,Doyon}  to the gapless case and show that, in the rather simple case of the resonant level model, and after some additional renormalizations, the form factors approach  yields  remarkably accurate  results for the entanglement  all the way from short to large distances. This is confirmed by detailed comparison with numerical simulations. Both our form factor and numerical results are compatible with a non-perturbative form at short distance.

\end{abstract}


\maketitle

\section{Introduction}

Quantum entanglement has given rise to much work in the condensed matter community as a new way to explore interesting aspects of physical systems. The Kondo problem for instance has been revisited along these lines, with studies addressing the interplay between the impurity screening and the information shared between the impurity and the bath~\cite{AffleckLaflorencie1,AffleckLaflorencie2}. It is certainly reasonable to expect that entanglement -- together with other quantities inspired by quantum information theory, such as the Loschmidt echo or the work distribution -- might shed new light on, and offer new experimental/numerical probes of, the key physical features of the Kondo and other problems~\cite{ChoKenzie,AffleckLaflorencie1}. A particularly interesting question in this direction is whether the Kondo screening cloud -- which has had  so elusive an appearance in standard thermodynamic quantities~\cite{Affleck} -- might play a bigger role in quantum information aspects. Other aspects of interest in 
the context of 2 level systems interacting with gapless excitations -- generalizing the Kondo problem -- apply to the decoherence of qubits interacting with the environment~\cite{LeHur1,LeHur2,LeHur3}. 

A large part of the  work combining entanglement and quantum impurities has been numerical so far. Indeed, apart from the scale invariant  situations, where conformal invariance techniques have led to spectacular progress~\cite{CardyCalabrese1,CardyCalabrese2}, the general situations involving crossover are very difficult to tackle. This is mostly because the entanglement is a different kind of quantity, not amenable to simple Bethe ansatz calculations, for instance. There is, however, another reason for the relative lack of analytical results in this area: entanglement, being a zero temperature quantity, is naturally plagued by IR divergences, which make it non perturbative in the impurity strength. In that respect, it does behave somehow like  some properties of the Kondo screening cloud studied in~\cite{AffleckBarzykin,Affleck}. 

In order to clarify the main features of entanglement in the presence of impurities -- in particular its scaling properties, and flow from small to strong coupling -- we focus in this paper on a couple of representative situations, which we handle by a mix of analytical and numerical techniques. The lessons learned will be put to use in  forthcoming papers, with applications of more direct physical interest.

The paper is organized as follows. In section~\ref{sec:models} we discuss the basic models we want to study, and define precisely the entanglement entropy. In section~\ref{sec:UVpert} we put together the perturbative calculation of the entanglement at small coupling, and show that it is plagued by strong IR divergences. In section~\ref{sec:IRdiv} we discuss this difficulty in a more general context. In section~\ref{sec:smallcoup} we show how the non perturbative nature of the entanglement entropy can be obtained using general conformal field theoretic arguments. In section~\ref{sec:largecoup} we recall the principles of the large coupling expansion proposed in~\cite{AffleckLaflorencie2} and carried out to high order in~\cite{FretonIR}. When the dimension of the perturbation is $h=\frac{1}{2}$, we develop in section~\ref{sec:FF} the form-factor approach using the results of~\cite{Doyon,Doyon1}, and obtain non perturbative approximations for the entanglement extrapolating all the way from the UV to the IR limit. Finally, in section~\ref{sec:numerics} we compare our results with those of exact numerical calculations on large spin chains. The conclusion contains some last comments and prospect for future work. Finally, appendix~\ref{sec:relsIsing} contains a discussion of the equivalence between our impurity models when the dimension of the perturbation is $h=\frac{1}{2}$ to the boundary Ising model with a boundary magnetic field at special values of the coupling.

\section{Models and questions}\label{sec:models}

The main problem  we  study  in this paper -- though it has various, mathematically equivalent formulations, see below  -- is  the calculation of the entanglement of a region of length $2L$ centered on an `impurity' in an otherwise one dimensional, gapless quantum system. We characterize this entanglement by the von Neumann entropy $S = - {\rm Tr} \rho \ln \rho$, where $\rho$ is the reduced density matrix that has been formed by tracing over the degrees of freedom outside of the segment of length $2L$. 

An example of this setup is obtained by taking  two semi infinite XXZ chains  coupled by a weak link:
\begin{equation}
H=\sum_{-\infty}^{-1} J\left[S_i^xS_{i+1}^x+S_i^yS_{i+1}^y+\Delta S_i^zS_{i+1}^z\right]
+\sum_{1}^\infty  J\left[S_i^xS_{i+1}^x+S_i^yS_{i+1}^y+\Delta S_i^zS_{i+1}^z\right]+J'\left[S_0^xS_{1}^x+S_0^yS_{1}^y+\Delta S_0^zS_{1}^z\right]\label{firstH}.
\end{equation}
The bulk chains  are in a gapless Luttinger liquid phase for $-1<\Delta\leq 1$.  We shall consider  the case of anisotropy  $\Delta<0$, where the tunneling between the two half infinite chains  is  a relevant perturbation, and one observes healing at large scales. The case $\Delta=0$ is exactly marginal. We  parametrize  $\Delta=-\cos{\mu\pi\over 2}$, $\mu\in \left[0,1\right]$. We focus on the physics at energies much smaller than the band-width, where field theoretic results can be applied. 

Consider  then the entanglement of a region of length $2L$ centered on the modified link. We can easily surmise what this entanglement will look like in the high and low energy limits from the existing literature. Indeed, at  high energy, the system is effectively cut in half. Using the well known formula for the entropy of a region {\sl on the edge} of a conformal invariant system we have 
\begin{equation}
S_{\rm UV}=2\times \left[{1\over 6}\ln {2L\over \epsilon}+{s_1\over 2}+\ln g\right]={1\over 3}\ln {2L\over \epsilon}+s_1+\ln g_{\rm UV},
\end{equation}
where we used that the central charge is unity, $\epsilon$ is a UV cutoff of the order of the lattice spacing $a$,  and $\ln g_{\rm UV} = 2 \ln g$ where $\ln g$ is the boundary entropy~\cite{AffleckLudwig} associated with the conformal boundary condition corresponding to an open XXZ spin chain. The remaining constant term $s_1$ is non-universal as it obviously depends on the definition of the cutoff $\epsilon$.

On the other hand, at low energy, healing has taken place, the system behaves just as one ordinary quantum spin chain, and the entropy obeys the general form for a region of length $2L$ in the bulk of a conformal invariant system:
\begin{equation}
S_{\rm IR}={1\over 3}\ln {2L\over \epsilon}+s_1 + \ln g_{\rm IR},
\end{equation}
Here,  we have allowed for a term $\ln g_{\rm IR}$, which can be thought of as a residual contribution of the weak link at low energy. In general, comparing entanglements for bulk and boundary theories is indeed difficult, since the dependency of the cutoff $\epsilon$ on the physical cutoff (the lattice spacing in the spin chain $a$, which is the same in both geometries) is not universal, and not necessarily the same in the bulk and boundary cases. This important aspects is discussed in detail in \cite{Doyon}, see in particular section 6.2.1 in that reference. The point for us is that  the quantity $\ln g_{\rm UV} - \ln g_{\rm IR}$ is well defined, and  its value $\ln g_{\rm UV} - \ln g_{\rm IR}  = - \frac{1}{2} \ln \mu$ can be easily obtained from the folded version of the system (see eq.~(\ref{hamilbdr}) below). 

More generally, since  the bulk behavior of the entanglement entropy is not modified, it is natural to expect the existence of a scaling relation
\begin{equation}
S(L)-S_{\rm IR}= {\cal S}_{\rm imp}(LT_B)  \label{scaling},
\end{equation}
where the crossover scale, $T_B$, is expected to be  related to the coupling $J'$. ${\cal S}_{\rm imp}$ should be a monotonic function extrapolating between $- \frac{1}{2}\ln(\mu)$ at small values of the argument and $0$ at large values.

To proceed, and conveniently describe the field theory limit \cite{AffleckEggert},  we first observe that the problem, at low energy,  can be turned into a purely chiral one. Indeed, in the low energy limit, each half chain is equivalent to a combination of L and R moving excitations, and we formally map via a canonical transformation  the L moving sector into a R moving one so as to have two chiral `wires' representing the two half chains. The additional tunneling between the two chains becomes, in this language, a hopping term between two chiral wires. Bosonizing, forming odd and even combinations of the bosons for each wire, one finds that   the odd combination decouples, while for the even one obtains the simple hamiltonian, 
\begin{equation}
\label{hamil}
H=\int_{-\infty}^\infty {\rm d}x \
(\partial_x\phi_R)^2+
\lambda \cos\beta\phi_R(0).
\end{equation}
where ${\beta^2\over 8\pi}=\mu \equiv h$ is the conformal weight of the perturbation, and  we  have set the Fermi velocity $v_F=1$.  The dimension of the perturbation being $[{\rm length}]^{-\mu}$, we see that $T_B\propto \lambda^{1/(1-\mu)}\propto (J')^{1/(1-\mu)}$. 
One can also fold back this problem   into the  boundary   sine-Gordon model (BSG) with  Hamiltonian
\begin{equation}
H_{\rm BSG}=\int_{-\infty}^0 {\rm d}x 
{1\over 2}\left[(\partial_x\Phi)^2+\Pi^2\right]
+\lambda \cos{\beta\over 2}\Phi(0)\label{hamilbdr}.
\end{equation}
This shows equivalence to a large variety of other problems, including the one of  tunneling between edge states in the Fractional Quantum Hall Effect (FQHE)~\cite{WenFQHE,ConductanceFQHE}.  In this case, $\mu=\nu$ is the filling fraction. The RG flows from Neumann ($\lambda=0$) to Dirichlet ($\lambda=\infty$) boundary conditions (BC), and the boundary entropy associated with these conformally invariant BC satisfy $\ln g_{\rm UV} - \ln g_{\rm IR} = - \frac{1}{2} \ln \mu$, as claimed earlier.

An interesting variant involves modifying two successive links on the chain:
\begin{eqnarray}
H&=&\sum_{-\infty}^{-2} J\left[S_i^xS_{i+1}^x+S_i^yS_{i+1}^y+\Delta S_i^zS_{i+1}^z\right]
+\sum_{1}^\infty  J\left[S_i^xS_{i+1}^x+S_i^yS_{i+1}^y+\Delta S_i^zS_{i+1}^z\right]+\nonumber\\
&+& J'\left[S_{-1}^xS_{0}^x+S_{-1}^yS_{0}^y+\Delta S_{-1}^zS_{0}^z+S_0^xS_{1}^x+S_0^yS_{1}^y+\Delta S_0^zS_{1}^z\right].
\end{eqnarray}
This is equivalent to  tunneling  through a resonant level at the origin. This time,  the dimension of the tunneling operator is half what it is in the previous situation, $J'$ is always relevant for $-1<\Delta\leq 1$, and the system is always healed at low energy.  The same series of manipulations --  `unfolding the two half chains', forming odd and even combinations, decoupling the odd one and bosonizing -- lead to the hamiltonian formulation 
\begin{equation}
H=\int_{-\infty}^\infty {\rm d}x \left(\partial_x\phi_R\right)^2+\lambda \left[{\mathrm e}^{i{{\beta\over\sqrt{2}}}\phi_R(0)}S^-+{\mathrm e}^{-i{{\beta\over\sqrt{2}}}\phi_R(0)}S^+\right],\label{chiralK}
\end{equation}
where we recall that  ${\beta^2\over 8\pi}=\mu$ ({\bf Post-publication footnote added}~\footnote{Minor erratum: unfortunately, the field theory~(\ref{chiralK}) does not correspond to the XXZ spin chain with two successive weak links as claimed here. The bosonization of the XXZ chain with two weak-links leads instead  to a field theory with {\it two} chiral bosons (one for each half-chain) tunneling through a spin-$\frac{1}{2}$ impurity. The dimension of the perturbation is also $h=\frac{\mu}{2}$ and the predictions of this paper can be applied to that case as well. The anisotropic Kondo Hamiltonian~(\ref{chiralK}) can be realized on the lattice as an interacting resonant level model (IRLM), with two half-infinite chains of non-interacting fermions tunneling through a resonant level model with Coulomb interactions on the dot only. In the XXZ language, this corresponds to an XX spin chain with two weak links, with anisotropy $\Delta \neq 0$ on the two weak links. All the conclusions of this paper remain unchanged. }) . In this case, the dimension of the perturbation is $h=\frac{\mu}{2}$ (note the factor ${1\over 2}$ compared with the first case). The problem can also be folded back into the anisotropic Kondo problem
\begin{equation}
H_{\rm AK}=\int_{-\infty}^0 {\rm d}x {1\over 2}\left[\left(\partial_x\Phi\right)^2+\Pi^2\right]+\lambda \left[{\mathrm e}^{i{\beta\over2\sqrt{2}}\Phi(0)}S^-+{\mathrm e}^{-i{\beta\over2\sqrt{2}}\Phi(0)}S^+\right].\label{bdrK}
\end{equation}

Of particular interest is the case   $\Delta=0$, $\mu=1$ which corresponds to free fermions. While the chain with one weak link is marginal, the chain with two weak links describes an interesting flow, and is in fact equivalent to a widely studied problem -- that of the resonant level model (RLM). Indeed, fermionization in this case leads to   %
\begin{equation}\label{eqRLMlattice}
H=-J\left(\sum_{-\infty}^{-2} c_{m+1}^\dagger c_m+\hbox{h.c.}\right)-J\left(\sum_{1}^{\infty} c_{m+1}^\dagger c_m+\hbox{h.c.}\right)-J'(c_{-1}^\dagger c_0+c_0^\dagger c_{-1}+c_0^\dagger c_1+c_1^\dagger c_0),
\end{equation}
where we have redefined the couplings $J \to -2J$ and $J' \to -2J'$. When going to the continuum limit, the $i=0$ site behaves like a two level impurity, and  the hamiltonian reads
\begin{eqnarray}
H=\int_{-\infty}^0 i\left[\psi_{1L}^\dagger\partial_x\psi_{1L}-\psi_{1R}^\dagger\partial_x\psi_{1R}\right] {\rm d} x+
\int_{0}^\infty i\left[\psi_{2L}^\dagger\partial_x\psi_{2L}-
\psi_{2R}^\dagger\partial_x\psi_{2R}\right] {\rm d} x+ \lambda \left[
\left(\psi^\dagger_{1}(0)+\psi^\dagger_2(0) \right)d+ {\rm h.c.} \right],\label{RLMferm}
\end{eqnarray}
with $\psi_{1L}(0)=\psi_{1R}(0)\equiv \psi_1(0)$, same for the second species, $\lambda\propto J'$\footnote{Having the Fermi velocity $v_F=1$, corresponds here to  $J=\frac{1}{2}$ in eq~(\ref{eqRLMlattice}).}. In contrast with the case of the XX chain with a single defect, this non-interacting problem  {\sl not scale invariant}. The coupling $\lambda$ flows, and the system again exhibits {\sl healing}: at low energy, the impurity level is completely hybridized with the two half chains.

Let us go back to the general case $\Delta=-\cos{\pi \mu\over 2}$. Proceeding like before, we can write the limiting behaviors of the entanglement entropy. At low energy, the impurity is hybridized, the system behaves just as {\sl one} non chiral wire and a hybridized impurity, and the entropy obeys the general form for a region of length $2L$ in the bulk of a conformal invariant system decoupled from the two baths, so 
\begin{equation}
S_{\rm IR}={1\over 3}\ln {2L\over \epsilon} + s_1 + \ln g_{\rm IR},
\end{equation}
where once again, we included a term $ g_{\rm IR}$ that accounts for the remaining boundary condition at $x=0$ of the hybridized impurity.
Meanwhile, at high energy, the impurity is completely decoupled from the wires, and one gets
\begin{equation}
S_{\rm UV}=2\times \left[{1\over 6}\ln {2L\over \epsilon}+{s_1\over 2}+\ln g\right]={1\over 3}\ln {2L\over \epsilon}+s_1+ \ln g_{\rm UV}.
\end{equation}
Using the folded (boundary) version of the system~(\ref{bdrK}), one can easily argue that $\ln g_{\rm UV} - \ln g_{\rm IR} = \ln 2$, as a decoupled impurity has two degrees of freedom. One thus expects a behavior entirely similar to~(\ref{scaling}), where the crossover scale, $T_B$, is expected to be proportional to a power of  the coupling square, $T_B\propto \lambda^{2/(2-\mu)}$, and ${\cal S}_{\rm imp}$ should be a monotonic function extrapolating between $\ln 2$ at small values of the argument and $0$ at large values.

Finally, we note that in the boundary versions (\ref{hamilbdr},\ref{bdrK}), the entanglement impurity we have discussed is now the entanglement of a region of length $L$ on the edge of the system with the rest. If one were to start from an (anisotropic) Kondo version, this would be the most natural point of view~\cite{AffleckLaflorencie1}.

There are of course other variants of the problem, for instance involving a slightly modified  link in the antiferromagnetic  XXZ chain with $\Delta>0$, interactions in the RLM model, etc. In all these cases, we should stress that the geometry we are considering is probably not the most interesting:   considering the entanglement of the two halves connected by a weak link or a quantum dot is probably  more physical. This latter problem is however significantly  more difficult technically. We will discuss it in our next paper, relying on the present work as a stepping stone. 

\section{UV perturbation}\label{sec:UVpert}

The most natural to explore the behavior of ${\cal S}_{\rm imp}(LT_B)$ between the fixed points is to use perturbation theory. The required calculation is a modification of the one proposed in~\cite{Holzhey,CardyCalabrese1,CardyCalabrese2}. Using the well-known replica trick, one first observes that the entanglement entropy $S$ can be obtained  from the Renyi entropies $R_n={\rm Tr} \rho^n$ by considering  $S=- \lim_{n \to 1} {\partial \over \partial n} R_n$.The Renyi entropies in turn  can  be obtained  as $R_n=\frac{Z_n}{(Z_1)^n}$, where $Z_n$ is the partition function on a $n$-sheeted Riemann surface ${\cal R}_{n,1}$ with the sheets joined at a cut corresponding to the segment of length $2L$. The difference between the problem at hand and the conformal case is that now there is a perturbation inserted at the origin in the Hamiltonian formulation, which corresponds to the insertion of a perturbation along an imaginary time line for each of the  $n$ sheets. 
The modified partition functions $Z_n$ can in principle be expanded in powers of the coupling constant, and perturbative corrections  to the Renyi entropies and the entanglement entropies finally obtained. 
 
 To see what happens in more detail, we consider first hamiltonian~(\ref{hamil}). We start with 
$n=2$ sheets, and  write formally  the Renyi entropy as a functional integral for a pair of chiral bosons as
\begin{eqnarray}
R_2=\frac{Z_2}{(Z_1)^2}={\int_{\rm twist} [{\cal D} \phi_1][{\cal D}\phi_2]
\exp\left\{-A[\phi_1]-A[\phi_2]-
\lambda\int_{-\infty}^\infty\left[V_1(x=0,y)+V_2(x=0,y)\right]{\rm d}y\right\}\over
\left(\int [{\cal D}\phi]\exp\left\{-A[\phi]-\lambda\int_{-\infty}^\infty  V(x=0,y){\rm d}y\right\}\right)^2}\label{def},
\end{eqnarray}
where $V_i=\cos\beta\phi_i$ is the perturbation, and $A[\phi] = \int {\rm d}^2 x \mathcal{L} [\phi]$ is the free action. Note we are working in the chiral version, but have suppressed the `R' label in the fields for simplicity of notation. 
Finally, the label twist means the functional integral is evaluated with conditions around the cut
\begin{equation}
{\begin{array}{c}
\phi_1(-L \leq x\leq L,y=0^+)=\phi_2(-L\leq x\leq L,y=0^-),\\
\phi_2(-L \leq x\leq L,y=0^+)=\phi_1(-L\leq x\leq L,y=0^-). \end{array}}\label{eqTwist}
\end{equation}

If $\lambda=0$, the ratio (\ref{def}) is nothing but the correlation function of an (order two) twist operator corresponding to~(\ref{eqTwist}), which we will write then~\cite{Holzhey,CardyCalabrese1}
\begin{equation}
R_2(\lambda=0)={z_2\over z_1^2}=\langle \tau_2(L,0)\tilde{\tau}_2(- L,0)\rangle_{{\cal L}^{(2)},\mathbb{C}}\propto L^{-1/8}\label{easy},
\end{equation}
where $\langle \dots \rangle_{{\cal L}^{(2)},\mathbb{C}}$ means that the correlator is to be evaluated in the plane $\mathbb{C} = \mathbb{R}^2$ (worldsheet) with the Lagrangian ${\cal L}^{(2)} = {\cal L} [\phi_1] + {\cal L} [\phi_2]$. We also recall that in general, the scaling dimension of the twist operator $\tau_n$ reads $h_n=\frac{c}{24}(n-1/n)$. We now consider the perturbation expansion in powers of $\lambda$. For the denominator, we have immediately
\begin{equation}
D=d^2;~~~d= z_1\left[1+{\lambda^2\over 2}\int {\rm d}y{\rm d}y' {1\over 2 |y-y'|^{2\mu}} + \dots \right]
\end{equation}
where the factor $1/2$ in the integral comes from the $1/2$'s in the cosines, and we recall that $\mu$ is the conformal weight $\mu={\beta^2\over 8\pi}$ of the perturbation. 

For the numerator, things are a little more complicated since we have two types of fields on the plane, with $1-1,1-2$ and $2-2$ contractions, in the presence of the glueing conditions along the cut. To proceed, we uniformize. We start with the complex coordinates $w=x + i y$, and introduce 
\begin{equation}\label{eqConfMap}
z=\left({w-u\over w-v}\right)^{1/2},
\end{equation}
where $u=L$ and $v=-L$ are the complex coordinates of the cut's extremities. This maps the whole 2-sheeted Riemann surface ${\cal R}_{2,1}$ to the $z$-complex plane $\mathbb{C}$.
%
%
We then write~(\ref{def}) as
\begin{equation}
{R_2\over R_2(\lambda=0)}={1+{\lambda^2\over 2}\int {\rm d}w{\rm d}w'\langle \cos\beta\phi(w)\cos\beta\phi(w')\rangle_{{\cal R}_{2,1}}
\over \left(1+{\lambda^2\over 2}\int {\rm d}y{\rm d}y' {1\over 2 |y-y'|^{2\mu}}\right)^2} + \dots \label{firstorderUV}
\end{equation}
where the spatial integrals in the numerator are now over ${\cal R}_{2,1}$ (worldsheet), and we have a unique boson $\phi$ instead of $\phi_1$ and $\phi_2$. Here the integrals in the numerator correspond to insertions along {\sl two} lines, corresponding to the two copies of the theory, so overall there are four  possible terms (contractions). The perturbation $V=\cos\beta\phi$  is a primary operator, so we can calculate the correlations on ${\cal R}_{2,1}$ by using the conformal mapping~(\ref{eqConfMap}). We have
\begin{equation}
2\times 2^{2\mu}\langle  \cos\beta\phi(w)\cos\beta\phi(w')\rangle_{{\cal R}_{2,1}}={(u-v)^{2\mu}\over 
N_{12}^{2\mu}(w-u)^{\mu/2}(w-v)^{\mu/2}(w'-u)^{\mu/2}(w'-v)^{\mu/2}},
\end{equation}
Here, 
\begin{equation}
N_{12}=(w-u)^{1/2}(w'-v)^{1/2}-(w'-u)^{1/2}(w-v)^{1/2}.
\end{equation}
In (\ref{firstorderUV}) we have to integrate $w,w'$ both over the imaginary axis $w=iy$, but also over the second sheet, which is obtained by sending $(w-u)\to {\mathrm e}^{2i\pi}(w-u)$ and same for $w'$. This means we end up with two integrals where $w,w'$ are on the same sheet, and two where they are on different sheets. 

Replacing everything by the particular choice of coordinates, and expanding the denominator in (\ref{firstorderUV}) we get
\begin{equation}
{R_2\over R_2(\lambda=0)}=1+{\lambda^2\over 4}\times 2\int_{-\infty}^\infty {\rm d}y{\rm d}y' \left[G_{\rm same}(y,y')+G_{\rm diff}(y,y')\right] + \dots,
\end{equation}
where the factor $2$ comes since there are two sheets, and insertions can be on same or different sheets, so
\begin{equation}
G_{\rm same}(y,y')={1\over 2^{2\mu}|y-y'|^{2\mu}}
{\left[(iy-L)^{1/2}(iy'+L)^{1/2}+(iy'-L)^{1/2}(iy+L)^{1/2}\right]^{2\mu}\over (y^2+L^2)^{\mu/2}((y')^2+L^2)^{\mu/2}}- \frac{1}{|y-y'|^{2\mu}},
\end{equation}
while $G_{\rm diff}$ will be the same expression with a minus in the numerator's bracket, and no subtraction (the two point function of the fields in different copies in the denominator of course vanish identically).

It is convenient to introduce new variables via $\tan\theta={y\over L}$, so the integral becomes
\begin{equation}
 L^{2-2\mu}\int_{-\pi/2}^{\pi/2} {\rm d}\theta {\rm d}\theta' (\cos^2\theta)^{\mu-1}(\cos^2\theta')^{\mu-1}
{(\cos^2{\theta-\theta'\over 2})^\mu-1\over(\sin^2(\theta-\theta'))^\mu}
\end{equation}
The second integral reads similarly
\begin{equation}
 L^{2-2\mu}\int_{-\pi/2}^{\pi/2} {\rm d}\theta {\rm d}\theta' (\cos^2\theta)^{\mu-1}(\cos^2\theta')^{\mu-1}
{1\over(\cos^2{\theta-\theta'\over 2})^\mu}
\end{equation}
Both integrals are UV convergent for a relevant perturbation $\mu<1$. They are however both IR divergent (here the IR region being $\theta\approx \pm{\pi\over 2}$).  This means that, although formally the perturbation at small coupling looks like it should be an expansion in powers of $\lambda^2 L^{2-2\mu}$, this might actually not be the case and, as we shall see later, is not. In fact, we will see that the entanglement is simply non perturbative in $\lambda^2$, and cannot be obtained via this perturbation theory. 

\smallskip

This result could appear as a surprise. On the one hand, the entanglement is a $T=0$ quantity, and such quantities are often plagued by IR divergences. On the other hand, we are looking for an $L$ dependent quantity, and it would be natural to expect that $L$ would act as an effective IR cutoff, rendering the perturbation expansion finite. This is however definitely not what happens. The situation is reminiscent of similar divergences encountered in the Kondo screening cloud problem \cite{Affleck}.

We stress finally that the argument applies almost without modification to the Hamiltonian (\ref{chiralK}). All that changes is that the perturbation is of the form ${\mathrm e}^{i\beta\phi}S^-+ {\rm h.c.}$ instead of ${\mathrm e}^{i\beta\phi}+{\rm h.c.}$, so exponentials of opposite signs have to alternate in the imaginary time insertions, modifying some of the numerical coefficients, but not the integrals or their divergences.

\section{IR divergences in quantum impurity problems}\label{sec:IRdiv}

To gain a better understanding of the situation, it is useful to  start by discussing  another observable~\footnote{This section somewhat lies outside of the main flow of this paper, as it does not deal directly with the computation of the entanglement entropy. It does contain however some very important points for our purpose, but the reader interested only in entanglement may wish to skip this section.} than the entanglement entropy for (\ref{hamil}). We turn briefly to the boundary formulation (\ref{hamilbdr}), and consider the one point function $\langle \cos{\beta\Phi\over 2}(x)\rangle$ which appears, for instance, in the determination of Friedel oscillations for impurities in Luttinger liquids. Simple scaling arguments suggest the general form
\begin{equation}
\langle :\cos{\beta\Phi\over 2}:\rangle =
\left(\frac{2}{|x|}\right)^{\mu/2}F\left(\lambda  |x|^{1-\mu}\right),
\label{scalingfct}
\end{equation}
where $F$ is a universal function obeying $F(\infty)=1$, so the field sees Dirichlet boundary conditions, and the bulk normalization has been chosen appropriately. 

Determining the function $F$ is also  a difficult problem. The most natural is once again to attempt perturbation theory in $\lambda$.  This would share many features of the calculation of one and two  point functions in the bulk sine-Gordon theory \cite{KL}. There, it is well known that 
 (in fact, the result essentially goes back to Coleman), 
provided $h\leq 1$ (that is, the perturbing operator is not 
irrelevant), there are {\sl no UV divergences} in the calculation. 
All divergences coming from bringing together two insertions of the 
perturbing term $\cos{\beta\Phi\over 2}$ are exactly canceled 
by similar divergences coming from the expansion of the denominator 
(the partition function and associated bubble diagrams). In general, 
the divergences are indeed controlled by the Operator Product Expansion (OPE), 
$$
{\mathrm e}^{i\beta\Phi(y)/2}{\mathrm e}^{-i\beta\Phi(y')/2}=|y-y'|^{-2\mu}
\left(1+\ldots-\pi \mu 
(y-y')^{2}(\partial \Phi(y))^{2}+\ldots\right),
$$
with all fields at $x=0$, $y$ the coordinate along the boundary.
The leading order comes from the contribution of the identity operator 
and leads to a disconnected piece subtracted off by a similar term in 
the denominator. The $\dots$ stand for higher orders, or lower orders 
that vanish after integration. The overall singularity at order 
$O(\lambda^{2n-1})$ thus behaves as $\sim \int \prod_{i=0}^{n-1} {\rm d} (y_{2 i} - y_{2 i+1}) \times (y_{2 i} - y_{2 i+1})^{2-2 \mu}$, it comes with dimension $d_{\rm UV}=n(3-2\mu)$ and 
for $\mu\leq 1$ the integrals are UV finite. Other  singularities 
(when several points are brought together at once etc \cite{KL}) behave 
similarly.

However, there will 
always appear IR divergences at a certain order, depending on the 
exact value of the conformal weight $h$. Of course, we expect in the end the scaling form 
 to hold, and thus to depend only on 
$xT_{B}$, $T_{B}\propto \lambda^{1/1-\mu}$. What will happen in general 
is that the divergences in the perturbative expansion have to be 
resummed before the proper scaling form can be obtained. The latter, 
in general, will thus behave non perturbatively in the coupling $\lambda$.

This is nicely illustrated in the case $\mu={1\over 2}$, where the 
the exact form  of the one point function  is known, thanks to a mapping to the boundary Ising model (see below), together with a very clever argument by Chatterjee and Zamolodchikov \cite{CZ}. One finds  \cite{LLS}
 \begin{equation}
\langle :\cos{\beta\Phi\over 2}:\rangle (x)=4 \lambda\sqrt{\pi} \left({x\over 2}\right)^{1/4}\Psi(1/2,1;8\pi \lambda^2 x)
\end{equation}
where $\Psi$ is the degenerate hypergeometric function. The asymptotics follows from $\Psi(1/2,1;2x)={{\mathrm e}^x\over\sqrt{\pi}} K_0(x)$, where $K_0$ is the usual modified Bessel function, 
%
%
%
so that we find
\begin{eqnarray}
\langle :\cos{\beta\Phi\over 2}:\rangle(x)&\sim &\left(\frac{2}{x}\right)^{1/4},~x \gg 1,\nonumber\\
\langle :\cos{\beta\Phi\over 2}:\rangle(x)&\sim & x^{-1/4} 2^{7/4} \lambda x^{1/2}\times -\ln(\lambda^2 x),~x \ll 1.
\end{eqnarray}
We thus see that this function exhibits a non perturbative dependence at small coupling $\lambda$. The non analyticity  in $\lambda$  arises from the IR divergence of the first perturbative integral.
 
 There is a general way to understand the non analyticity of course. Whenever  a bulk operator (of conformal weights $h,h$) is sent to the boundary where it becomes a boundary field of weight $h_B$, one has
 \begin{equation}
 O(x)\approx x^{h_B-2h} O_B+\ldots
 \end{equation}
 In our case, the cosine of the bulk field simply goes over to the cosine of the boundary field. We have  thus $h={\mu\over 4}$ and $h_B=\mu$ , while $[O_B]=L^{-h_B}\propto T_B^{h_B}\propto\lambda^{h_B/1-\mu}$. We thus expect that 
 $\langle O_B\rangle=\langle :\cos{\beta\Phi(0)\over 2}:\rangle\propto \lambda^{\mu/1-\mu}$. The dependence of the one point function of the boundary field on $\lambda$ is non analytic in $\lambda$, and  non perturbative - of course, because again of IR divergences. This problem is the cousin of a similar problem in bulk massive theories, and has been studied in \cite{Zamo,FLZZ}. We deduce from this that, to leading order,
 \begin{equation}
\langle   O(x)\rangle \propto x^{h_B-2h} \lambda^{h_B/1-\mu}
 \end{equation}
More generally, we can write
\begin{equation}
 O(x)\approx \sum_i x^{h_B^i-2h} O^i_B
 \end{equation}
 so
\begin{equation}
 \langle O(x)\rangle \approx \sum_i x^{h_B^i-2h} c^i_B T_B^{h_B^i}=x^{-2h}\sum_i x^{h_B^i}c^i_BT_B^{h_B^i}\equiv x^{-2h} G(xT_B)
 \end{equation}
with $G(xT_B)\equiv F( \lambda x^{1-\mu})$
\begin{equation}
F(y)\propto y^{h_B}\propto \lambda^{h_B/1-\mu} x^{h_B}
\end{equation}
Going back to  the case of Friedel oscillations, we have therefore $F(y)\propto y^{\mu}\propto x^\mu \lambda^{\mu/1-\mu}$. This leading dependence in $\lambda$ replaces the expected perturbative one, which would be  linear in $\lambda$. 

The foregoing argument applies in the generic case. Whenever there are ``resonances'' and the parameter $\mu$ takes special rational values $\mu=1-{1\over 2n}$ , extra logarithmic terms appear in the one point functions of the operators right on the boundary, which translates in logarithms in the one point functions of operators at $x\neq 0$ as well. This is the case precisely when $\mu={1\over 2}$. 

It is important to stress also that the IR divergences naturally disappear at finite temperature, $1/T$ providing a natural cutoff. Once again this is illustrated in the $\mu={1\over 2}$ case, where one finds \cite{LLS,SM}, for Friedel oscillations at finite temperature
\begin{equation}
\langle :\cos{\beta\Phi\over 2}:\rangle(x)=f(2\lambda^2/T) \left({4\pi T\over\sinh(2\pi Tx)}\right)^{1/4} F\left({1\over 2},{1\over 2};1+2{\lambda^2\over T},{1-\coth 2\pi xT\over 2}\right)
\end{equation}
Here  $F$ is the usual hypergeometric function, $f$ is a function whose existence and value were determined in \cite{SM}. The right hand side admits a perturbative expansion in powers of $\lambda$, whose leading term, at fixed $x$, goes as $\lambda/\sqrt{T}$ when $T\to 0$. The coefficient of $\lambda$ thus  diverges in the zero temperature limit, in agreement with the fact that the true expansion is then in $\lambda\ln\lambda$. 

The general  IR behavior can easily be investigated. One finds that at order $O(\lambda^{2n+1})$, there is no IR divergence provided $\mu>{n+1/2\over n+1}$. Only when $\mu=1$ -- that is, the boundary perturbation is exactly marginal, and the bulk is a Fermi liquid --  are all orders finite. In this case, the Friedel oscillations admit a perturbative expansion in powers of $\lambda$~\cite{Egger}.


%
%
%
%

While the nature of the divergences is quite generic, the  quantities for which  they occur depend on the problem at hand. For instance, for the screening cloud in the (anisotropic) Kondo model, divergences occur even when the boundary perturbation has dimension one - in that case, it is marginally relevant~\cite{Affleck}. 

\section{The small coupling behavior of the entanglement entropy}\label{sec:smallcoup}

We now go back to the calculation of the entanglement entropy for hamiltonian~(\ref{hamil}). We see that, to obtain the non perturbative UV behavior, we must discuss twist fields and their OPEs. We follow the paper~\cite{CCT} but focus more directly on the question at hand. Imagine we have a single interval for which we want to calculate the entanglement with the rest of the system, and introduce accordingly the $n$-sheeted Riemann surface ($n$ replicas) ${\cal R}_{n,1}$. In the limit where the interval of length $L$ shrinks, we expect the presence of the two sewing points to decompose like an operator product expansion of the form
\begin{equation}
I=\sum_{\{k_j\}} C_{\{k_j\}} \prod_{j=1}^n \Psi_{k_j}(z_j)\label{opetwist}
\end{equation}
where we allowed for fields inserted at points $z_j$, the point $z$ on the $j^{th}$ sheet, and the set $\{\Psi_k\}$ denotes a complete set of local fields for one copy of the CFT. Recall that the cut in the Riemann surface ${\cal R}_{n,1}$ corresponds to the insertion of twist fields in the complex plane, so that $I \sim \tau_n(L)\tilde{\tau}_n(-L)$ and~(\ref{opetwist}) should be considered as the OPE of these twist fields. What~(\ref{opetwist}) means more precisely is that, if we have other operators inserted elsewhere, we can expect to have
\begin{equation}
{Z_n(L)\over Z_1^n}\langle \prod_{j=1}^n {\cal O}_j\rangle_{{\cal R}_{n,1}}=\langle I\prod_{j=1}^n {\cal O}_j\rangle_{\mathbb{C}^n}=\sum_{\{k_j\}} C_{\{k_j\}} \prod_{j=1}^n \langle \Psi_{k_j}(z_j){\cal O}_j\rangle_{\mathbb{C}_j}
\end{equation}
where $O_j$ designates operators inserted on the $j^{th}$ sheet, and $\mathbb{C}_j$ is the $j^{th}$ copy of the complex plane. Note indeed that the expectation on the right is taken in a fully factorized theory. 

Restricting now to the  $\Psi_k$  that make an orthonormal basis (so in particular they are all quasiprimary), and choosing $O_j=\Psi_{k_j}$ shows that the structure constant $C_{\{k_j\}}$ will not vanish only if the average of $\prod_j {\cal O}_j$ on the Riemann surface ${\cal R}_{n,1}$ does not vanish. It is useful to make things concrete now, so for instance we see that there is no term with a single primary operator on the right hand side of~(\ref{opetwist}) since the corresponding one point function on ${\cal R}_{n,1}$ vanishes. There is, however, at least one term with a single operator, the stress energy tensor, since we know that $\langle T\rangle_{{\cal R}_{n,1}}\neq 0$. Apart from this, the most important terms will be those involving the same primary operator on two different sheets $\prod_{j=1}^n\Psi_{k_j}=\Psi_{1}\Psi_{2}$, whose average on ${\cal R}_{n,1}$ will be non zero in general. If the  field $\Psi$ has conformal weights $h,\bar{h}$, we will thus have that
\begin{equation}
C\propto L^{-4h_n} L^{2\times(h+\bar{h})}
\end{equation}
where $h_n={c\over 24}\left(n-{1\over n}\right)$ is the conformal weight of the twist field. The crucial point is that $C$ involves {\sl twice} the scaling dimension of primary fields, in contrast with ordinary OPEs where only the scaling dimension would appear. 

The discussion carries over to the boundary case. One can, for instance, think of it after unfolding the system so as to keep only chiral fields as in~(\ref{hamil}). Everything then formally goes through after setting $\bar{h}=0$.  The question is then, what kind of fields $\psi$ (the chiral part of $\Psi$) can appear in the OPE of two twist fields. The one copy bulk theory is a compact boson which allows for the fields $\exp\left(\pm i{\beta\over 2}\Phi\right)$ on the boundary. This means that the radius is $R={2\over \beta}$, and thus the bulk conformal weights are given by 
\begin{equation}
\Delta_{wk}=2\pi\left({\beta k\over 8\pi}-{w\over \beta}\right)^2,~~\bar{\Delta}_{wk}=2\pi\left({\beta k\over 8\pi }+{w\over \beta}\right)^2
\end{equation}
Restricting to scalar operators we get $\Delta={k^2\beta^2\over 32\pi}$ or $\Delta={2\pi\over \beta^2}w^2$. 
For instance, the first values of $\Delta$ correspond to fields $\exp\left(\pm ik{\beta\over 2}\Phi\right)$, or, for the chiral part, 
$\exp\left(\pm ik{\beta\over 2}\phi_R\right)$. 

We now go back to the entropy calculation in the folded, non-chiral theory~(\ref{hamilbdr}). Upon folding, the chiral vertex operators $\mathrm{e}^{\pm ik{\beta\over 2}\phi_R(0)}$ become  $\mathrm{e}^{\pm ik{\beta\over 4}\Phi(0)}$, as $\Phi(0) = \phi_R(0)+\phi_L(0)=2 \phi_R(0)$. Recall also that the non-chiral twist field in the folded version can be thought of as the chiral part of $I$ in the unfolded theory. Hence, going through the discussion of short distance expansions we find, for the non chiral twist field
\begin{equation}
\tau_n(L)\approx L^{-2h_n}\left(1+\sum_k L^{2\Delta_k}c_n\sum_{\substack{ i,j=1\\\ i \neq j}}^n {\mathrm e}^{ik{\beta\over 4}\Phi_i(0)}{\mathrm e}^{ik{\beta\over 4}\Phi_j(0)}+\ldots\right)
\end{equation}
where we used the fact~\cite{CCT} that the two fields $\psi$ in the twist OPEs must belong to different copies. We are only interested in terms whose one point function acquires a non zero value in the presence of the perturbation. This means the first term with $k=1$ cannot contribute, and thus we need $k=2$, $\Delta_2=\mu$.  Taking derivative with respect to $n$ gives then the leading term for the entanglement correction, which should go as 
\begin{equation}
{\cal S}_{\rm imp} - \ln 2\propto (LT_B)^{2\mu}\propto L^{2\mu}\lambda^{2\mu/(1-\mu)}\label{res}
\end{equation}
For $\mu={1\over 2}$ in particular, this can be corrected by a resonance, and it is tempting to speculate then that one has
\begin{equation}
{\cal S}_{\rm imp}- \ln 2\propto LT_B\left[\hbox{cst}+\hbox{cst}\ln(LT_B)\right]\label{slog}
\end{equation}
Finally, we note once again that the RLM or the various (anisotropic) Kondo versions will behave identically, the presence of the operators $S^+,S^-$ not modifying in any essential way the OPE argument -- but one will have to be careful with the dimensions of the operators involved, and their relationship with $\mu$.  In the end, we find that for the RLM~(\ref{slog}) is expected  to hold as well.

\section{Large coupling expansion}\label{sec:largecoup}

While the small coupling expansion is plagued with IR divergences, a large coupling expansion is possible. It is now finite in the IR, and exhibits UV divergences which are easily taken care of using integrability and analyticity. Let us recall how the calculation goes at leading order in the anisotropic Kondo case~\cite{ImpEnt} (see also {\it e.g.}~\cite{Eriksson}). The leading IR perturbation is nothing but the stress energy tensor $H=H_{\rm IR}+\frac{1}{\pi T_B} T(0)+\dots$ The correction to the Renyi entropy can therefore be expressed as
\begin{equation}
-\delta Z_n = \frac{n}{\pi T_B} \int_{-\infty}^{+\infty} {\rm d} \tau \langle T(w=i \tau) \rangle_{{\cal R}_{n,1}}= \frac{n-n^{-1}}{24 \pi T_B} \int_{-\infty}^{+\infty}  \frac{(2L)^2}{(i \tau -L)^2(i \tau +L)^2} {\rm d} \tau= \frac{1}{12 L T_B} \left(n - \frac{1}{n} \right).
\end{equation}
The first correction to the entanglement entropy thus reads
\begin{equation}
{\cal S}_{\rm imp} = \frac{1}{6 L T_B} + \dots
\end{equation}
It is quite remarkable that this result does not depend on the anisotropy parameter $\frac{\mu}{2}$ (recall that $\Delta= - \cos \pi \frac{\mu}{2}$ in the XXZ language). It turns out that this IR expansion can be generalized to higher orders~\cite{FretonIR}. The results for the Kondo case are as follows
\begin{eqnarray}
{\cal S}_{\rm imp}&=&\frac{1}{6}\,\ln \left(1+\frac{1}{LT_B}\right)  - {18\over 35} {(\pi g_4)^2\over (2LT_B)^6}(4\alpha^4-8\alpha^2+9)+{\cal O}((LT_B)^{-7})
\label{result:infinite:imp}
\end{eqnarray}
where the coefficient $g_4$ has the following dependence on the dimension $h={\mu\over 2}$ of the tunneling operator 
\begin{equation}
g_4={\mu\over 12\pi^2} \left({\Gamma(\mu/2(2-\mu))\over \Gamma(1/(2-\mu))}\right)^3{\Gamma(3/(2-\mu))\over \Gamma(3\mu/2(2-\mu))},~~\alpha=\frac{(2-\mu)}{\sqrt{2\mu}}.
\end{equation}
Note that in (\ref{result:infinite:imp}), the first term in the right hand side  has  to be truncated at order 6. 

While in principle higher orders in the IR expansion could be determined, the complexity of the calculations   increases considerably. Moreover, the convergence properties of  this expansion are not clear. 
Finally, we observe that, in this point of view, the pure BSG case turns out to be quite different, because different operators appear in the IR effective description. The corresponding result has not even been worked out yet. 

Making analytical progress  therefore requires developing non perturbative approaches. The problems we are interested in are indeed integrable, at least in their boundary versions. While it is  natural to expect that this can be used in some way, integrability has been mostly used to calculate local properties such as magnetization, energy or impurity entropies. Von Neumann entanglement is non local, and therefore much harder to obtain in general.

\section{Form factor approach to the entanglement entropy }\label{sec:FF}

We will in what follows restrict to the case where the dimension of the perturbation is $h=\frac{1}{2}$: this corresponds to $\Delta=-{\sqrt{2}\over 2}$ ($\mu=\frac{1}{2}$) for the problem of tunneling between XXZ chains, and to $\Delta=0$ -- the RLM ($\mu=1$) -- for the tunneling through an impurity. These cases are closely related to the boundary Ising model with a boundary magnetic field (see Appendix~\ref{sec:relsIsing}). While the problem of calculating the entanglement non perturbatively remains extremely difficult -- entanglement still involving non local observables in the fermionic language -- it can be tackled using the idea of form-factors. 

 It has been known for many years that correlation functions of local observables in massive integrable theories can be calculated using the form-factors approach, where the integrable quasiparticles provide a basis of the Hilbert space, and the form-factors (FF) -- that is, the matrix elements of the operators in that basis -- can be obtained using an axiomatic approach based on the knowledge of the S matrix and the bootstrap. It is a natural idea to extend this approach to the case of entanglement entropy. Indeed, the Von Neumann entanglement is obtained form the Renyi entropy by taking an $n$ derivative at $n=1$, and the Renyi entropies can be considered formally as correlation functions of twist operators that live in $n$ copies of the theory of interest. The integrability of a single theory carries over to integrability of the $n$ copies, and a calculation similar to the one of ordinary correlators can be set up, after some additional work to determine the form factors of the twist operators $\tau,\tilde{\tau}$~\cite{Doyon,Doyon1}. 
 
 We are interested here in a variant where the bulk is massless. The form-factors technique in this case is more delicate to use, since particles can have arbitrarily low energies, and the convergence of the approach is not guaranteed. Various regularization tricks have to be used in the calculation of local quantities (eg the charge density for Friedel oscillations) \cite{Skorik, LS}, and we will see below that the situation for the entanglement is not better. Nevertheless, ${\cal S}_{\rm imp}(LT_B)$ can be calculated for $h=\frac{1}{2}$, by using the Ising model formulation, and relying heavily on the work~\cite{Doyon,Doyon1}.

To fix ideas, and explore the feasibility of form-factors calculations in our problem, we first discuss briefly the bulk case and the massless limit. One can find in~\cite{Doyon,Doyon1} the first order contribution to the two point function of the bulk Ising model twist field in the bulk 
\begin{equation}
\langle \tau(r)\tilde{\tau}(0)\rangle=\langle \tau\rangle^2+{1\over 2}\sum_{i,j=1}^n \int {{\rm d}\theta_1\over 2\pi}{{\rm d}\theta_2\over 2\pi}|F_2^{\tau|ij}(\theta_{12},n)|^2{\mathrm e}^{-mr(\cosh\theta_1+\cosh\theta_2)} + \ldots \label{firstorder}
\end{equation}
where $n$ is the number of copies, $\theta_i$ is the rapidity of the $i^{\rm th}$ particle with energy $e=m \cosh \theta_i$ and momentum $p=m \sinh \theta_i$, and $F_2^{\tau|ij}(\theta_{12},n)$ is the two-particle form factor of the twist field $\tau$
\begin{equation}
F_2^{\tau|ij}(\theta_{12},n) = \Braket{0 \left| \tau(0) Z^\dag_i(\theta_1)Z^\dag_j(\theta_2) \right|0}.
\end{equation}
In this last expression, we have used the notation $Z^\dag_j$ for the usual Faddeev-Zamolodchikov creation operators (here, the fermions) living in the $j^{\rm th}$ copy.
Since the theory is integrable, the form factors $F_2^{\tau|ij}(\theta_{12},n)$ can be computed exactly and are conveniently expressed using the function
\begin{equation}
K(\theta)={F_2^{\tau|11}\over \langle \tau\rangle}=-i{\cos{\pi\over 2n}\sinh{\theta\over 2n}\over n\sinh{i\pi+\theta\over 2n}\sinh{i\pi-\theta\over 2n}}
\end{equation}
which vanishes when $n=1$. The other form factors $F_2^{\tau|ij}(\theta_{12},n)$ can then be obtained from $F_2^{\tau|11}(\theta_{12},n)$ by shifting appropriately $\theta_{12}$ by a factor of $2 \pi i$.
Going to variables $\theta_1\pm \theta_2$ one can perform one integration, and be left with
\begin{equation}
\langle \tau(r)\tilde{\tau}(0)\rangle=\langle \tau\rangle^2\left(1+{n\over 4\pi} \int_{-\infty}^\infty {\rm d}\theta f(\theta,n)K_0(2mr\cosh(\theta/2))\right)+ \dots
\end{equation}
where 
\begin{equation}
\langle \tau\rangle^2f(\theta)\equiv |F_2^{\tau|11}(\theta)|^2+\sum_{j=1}^{n-1} |F_2^{\tau|11}(\theta+2i\pi j)|^2
\end{equation}
Doyon {\it et al.} then argue the crucial result that
\begin{equation}
\left.{{\rm d}\over {\rm d} n} nf(\theta,n)\right|_{n=1}={\pi^2\over 2}\delta(\theta)\label{deriv}.
\end{equation}
Taking the derivative of the two point twist correlation function meanwhile should give, at short distances, the entanglement entropy of the CFT. Since (\ref{firstorder}) is only a first order approximation where contributions with a larger number of particles have not been included, we get an approximation to the  entanglement entropy of a segment of length $r$ in the bulk with the rest of the system~\cite{Doyon1}
\begin{equation}
S_A=\ldots -{K_0(2mr)\over 8}+\ldots\approx \ldots+{\ln r\over 8}+\ldots
\end{equation}
and thus the expected  factor ${c\over 3}={1\over 6}$ is approximated by ${1\over 8}$  at this order.

Since in this paper we are interested in bulk CFTs, we need to take an  $m\to 0$ limit. This corresponds formally to describing the CFT  using massless  particles  and massless scattering. We thus set ${m\over 2}=M {\mathrm e}^{-\theta_0}$ and send $\theta_0\to \infty$. Only two types of excitations remain at finite energies: those for which $\theta=\pm\theta_0\pm \beta$ with $\beta$ finite. In the first case, one obtains right moving particles with $e=p=M {\mathrm e}^\beta$ and in the second case left moving particles with $e=-p=M{\mathrm e}^\beta$. Conformal fields factorizing into left and right components are not expected to mix the L and R sectors. Indeed, 
\begin{equation}
\hbox{lim}_{\theta\to\infty} K(\theta)=0,
\end{equation}
so only the LL and RR sectors will contribute in the massless limit of~(\ref{firstorder}). Therefore, setting (say for the R sector)
\begin{equation}
\theta_{1,2}=\theta_0+\beta_{1,2},
\end{equation}
and introducing $\beta_{\pm}\equiv \beta_1\pm \beta_2$ we obtain 
\begin{equation}
\langle \tau(r)\tilde{\tau}(0)\rangle=\langle \tau\rangle^2+{1\over 2}\sum_{i,j=1}^n \int {{\rm d}\beta_+\over 2\pi}{d\beta_-\over 2\pi}|F_2^{\tau|ij}(\beta_-,n)|^2{\mathrm e}^{-2M r {\mathrm e}^{\beta_+/2}\cosh(\beta_-/2)} + \dots
\end{equation}
where the $1/2$ coming from the Jacobian was canceled by the fact that there are two integrals, the L and the R one. Using (\ref{deriv}) we get the correction to the entanglement entropy as
\begin{equation}
S_A=\ldots -{1\over 16}\int_{-\infty}^\infty {\rm d}\beta_+{\mathrm e}^{-2M r {\mathrm e}^{\beta_+/2}}=\ldots-{1\over 8}\int_0^\infty {{\rm d}x\over x}{\mathrm e}^{-2M r x}\label{divint}
\end{equation}
This integral is divergent at small energy, a feature which is quite general in the use of massless form-factors.  We regularize by considering the integral
\begin{equation}
\int_0^\infty {\rm d}x x^{\alpha-1}{\mathrm e}^{-2M r x}={1\over (2M r)^\alpha} \Gamma(\alpha)=\left({1\over \alpha}+\ldots\right)\left(1-\alpha\ln(2M r)+\ldots\right)
\end{equation}
so the finite part of the integral is $-\ln(2M r)$ and thus we recover
\begin{equation}
S_A=\ldots+{1\over 8}\ln r+\ldots
\end{equation}

Let us  now consider the  Ising model with a  boundary magnetic field as in (\ref{Ising}), and to start assume that the bulk is massive.
The  form factors approach can be applied to this case as well. The first non trivial contribution  reads then~\cite{Doyon}
\begin{equation}
\langle 0|\tau(r)|B\rangle=\langle \tau\rangle+{1\over 2}\sum_{i=1}^n\int {{\rm d} \theta\over 2\pi}R\left({i\pi\over 2}-\theta\right){\mathrm e}^{-2mr\cosh\theta}F_2^{\tau|11}(-\theta,\theta,n)+\dots\label{firstord}
\end{equation}
coming from the boundary state
\begin{equation}
|B\rangle=\exp\left[{1\over 4\pi}\sum_{j=1}^n\int {\rm d} \theta R\left({i\pi\over 2}-\theta\right)Z^\dag_j(-\theta)Z^\dag_j(\theta)\right] \Ket{0},
\end{equation}
where we recall that $Z^\dag_j$ are the usual Faddeev-Zamolodchikov creation operators living in the $j^{\rm th}$ copy, and $R\left(\theta\right)$ is the reflection matrix~\cite{GZ} of the Ising field theory with a boundary magnetic field $h_b$ (proportional to $\lambda$ in \ref{Ising}). Ultimately, we want once again to take the massless limit $m \to 0$.
Notice that (\ref{firstord})  involves $F_2$ instead of $|F_2|^2$. We write
\begin{equation}
\langle 0|\tau(r)|B\rangle=\langle \tau\rangle+{n\over 4\pi}\int {\rm d} \theta R\left({i\pi\over 2}-\theta\right)F_2^{\tau|11}(-\theta,\theta){\mathrm e}^{-2mr\cosh\theta} + \dots
 \end{equation}
and observe that the analytical continuation in $n$ is trivial because the particle and its reflection must belong to the same copy. To every order, contributions are linear in $F$. But there is a lot of similarity -- eg between the term with four particles here, and the term with two particles in the bulk entropy. In the massless limit case, since the boundary produces as many L as R particles, and since we need both these numbers to be even, only the  terms with $2l$ R movers and  $2l$ L movers contribute. 

Indeed, the first correction to the entanglement, after taking derivative with respect to $n$ at $n=1$ reads explicitly 
\begin{equation}
s_1=-{1\over 4}\int_0^\infty {\rm d} \theta \left({\kappa+\cosh\theta\over \kappa-\cosh\theta}\right)\left({\cosh\theta-1\over \cosh^2\theta}\right){\mathrm e}^{-2mL\cosh\theta},
\end{equation}
with $\kappa = 1-h_b^2/(2 m)$. To obtain a scaling expression in the massless limit, we  boost rapidities like in the bulk case, 
and we obtain
\begin{equation}
s_1\approx {1\over 4}\int_{-\infty}^\infty {\rm d} \beta {{\mathrm e}^\beta-{h_b^2\over 2M}\over {\mathrm e}^\beta+{h_b^2\over 2M}}\times 2{\mathrm e}^{-\theta_0}{\mathrm e}^{-\beta} {\mathrm e}^{-2LM {\mathrm e}^\beta}\rightarrow 0,
\end{equation}
a vanishing result - natural, since in this limit, the two-particle form factor factorizes onto one particle form factors (one for the left, one for the right), which both vanish. We thus need to go to the next order (corresponding to 4 particles), where we use formula eq. (3.25) in \cite{Doyon}:
\begin{equation}
s_2={1\over 16}\int_0^\infty d\theta \left({\kappa+\cosh\theta\over \kappa-\cosh\theta}\right)^2\left({1-\cosh\theta\over 1+\cosh\theta}\right){\mathrm e}^{-4mL\cosh\theta}
\end{equation}
We obtain then 
\begin{eqnarray}
s_2\approx -{1\over 16}\int_{-\infty}^\infty {\rm d} \beta \left({{\mathrm e}^\beta-{h_b^2\over 2M}\over {\mathrm e}^\beta+{h_b^2\over 2M}}\right)^2 {\mathrm e}^{-4LM {\mathrm e}^\beta}\nonumber\\
= -{1\over 16}\int_{-\infty}^\infty {\rm d} \beta \left({e^\beta-T_B\over e^\beta+T_B}\right)^2 e^{-4L {\mathrm e}^\beta}\label{lowestorder}
\end{eqnarray}
where we have set
\begin{equation}
T_B\equiv {h_b^2\over 2}
\end{equation}
and we have shifted the $\beta$ integral. Now the expression (\ref{lowestorder}) is divergent at low energies, just like (\ref{divint}). To regularize it, we consider the difference:
\begin{equation}
s_2(LT_B)-s_2(\infty)={1\over 4} \int_{-\infty}^\infty {\rm d} \beta {{\mathrm e}^{\beta}\over (1+{\mathrm e}^{\beta})^2}{\mathrm e}^{-4LT_B {\mathrm e}^\beta}={1\over 4}\int_0^\infty {{\rm d} u\over (1+u)^2}{\mathrm e}^{-4LT_B u}
\end{equation}
 The UV value is ${1\over 4}=0.25$, to be compared with the exact value ${1\over 2}\ln2=0.346574\dots$. This indicates we are on the right track.

 To proceed, we now take equation (3.54) in \cite{Doyon}, perform the appropriate limits and rescalings to get
 \begin{eqnarray} \label{eqFFBOrder2}
 s_4\approx {1\over 2^8\pi^2}\int \prod_i {\rm d} \beta_i\delta(\sum\beta_i)
 \left[ {\mathrm e}^{-2L\sum {\mathrm e}^{\beta_i}} \prod_i {{\mathrm e}^{\beta_i}-T_B\over {\mathrm e}^{\beta_i}+T_B}\prod_i {1\over \cosh{\beta_i-\beta_{i+1}\over 2}}-\begin{array}
  {c}
  \beta_{1,3}\to \beta_{1,3} \pm {i\pi\over 4}\\
    \beta_{2,4}\to \beta_{2,4}\mp{i\pi\over 4}\end{array}\right]
    \end{eqnarray}
  where products and sums run over $i=1,\ldots,4$ and we have set $\beta_{4+1}\equiv \beta_1$.  The second term is obtained by shifting the contours of integration in the imaginary direction as indicated.  We observe the same divergence at low energy, and the same regularization (subtracting the formal expression for $s_4(\infty)$) also works like for $s_2$. We find
\begin{equation}
s_4(LT_B)-s_4(\infty)=\int_{0}^\infty {{\rm d} u_1 {\rm d} u_2 {\rm d} u_3 \over 16 \pi^2} \left[ \mathrm{e}^{-2 LT_B(u_1+u_2)(u_2+u_3)/u_2} \frac{u_2}{(u_1+u_2)^2(u_2+u_3)^2} \left( \prod_i \frac{1-u_i}{1+u_i} -1 \right) - \dots \right],
\end{equation}
where the dots correspond to the two other terms obtained by shifting the contours of integration as in~(\ref{eqFFBOrder2}). The UV value is  $s_4(0)-s_4(\infty)= \frac{1}{24}$, so at second order, we have the UV value $\frac{1}{4}+ \frac{1}{24} = 0.291667\dots $, to be compared once again with the value ${1\over 2}\ln2=0.346574\dots$ 

We find that higher orders can be dealt with in the same way, and that the UV values can be resummed exactly to yield the exact result
\begin{equation}
S_{\rm UV}-S_{\rm IR} = \sum_{l=1}^\infty \left( s_{2 l}(0)-s_{2 l}(\infty) \right) = \sum_{l=1}^\infty { 1 \over 4 l (2 l -1)} = {1\over 2}\ln 2, 
\end{equation}
as expected.

 We now return to  the RLM, whose results are obtained simply by multiplying those for  Ising by a factor of two. In the following, we will allow for an extra multiplicative renormalization to obtain the UV result exactly, that is consider, at lowest order,  the ratio 
\begin{equation}
\mathcal{S}^{(2)}_{\rm imp}(LT_B) \equiv \ln 2 {s_2(LT_B)-s_2(\infty)\over s_2(0)-s_2(\infty)}=\ln 2\int_0^\infty { {\rm d} u\over (1+u)^2} {\mathrm e}^{-4LT_B u}.
\end{equation}

It is then interesting  to consider the IR expansion of this quantity. Using
\begin{equation}
s_2(LT_B)-s_2(\infty)={1\over 2}\left[\alpha {\mathrm e}^\alpha {\rm Ei}(-\alpha)+1\right],~~~\alpha=4LT_B,
\end{equation}
where ${\rm Ei}$ is the usual exponential integral function. One finds 
\begin{equation}
\mathcal{S}^{(2)}_{\rm imp}(LT_B) = 2\ln 2\sum_{k=1}^n (-1)^{k-1} {k!\over 2(4LT_B)^k}+ \mathcal{O}\left(\frac{1}{(LT_B)^{n+1}} \right),
\end{equation}
where the expansion is only asymptotic. We see thus our `renormalized' first order approximation  interpolates between $\ln 2$ and  to ${\ln 2\over 4LT_B}=0.173287/(LT_B)$, while the exact result goes from $\ln 2$ to ${1\over 6LT_B}=0.16666/(LT_B)$, which is quite good. 

The next order approximation can be handled similarly, and we will simply provide the corresponding results on the curves below. 

\begin{figure}
\centering
    \includegraphics[scale=.75]
    {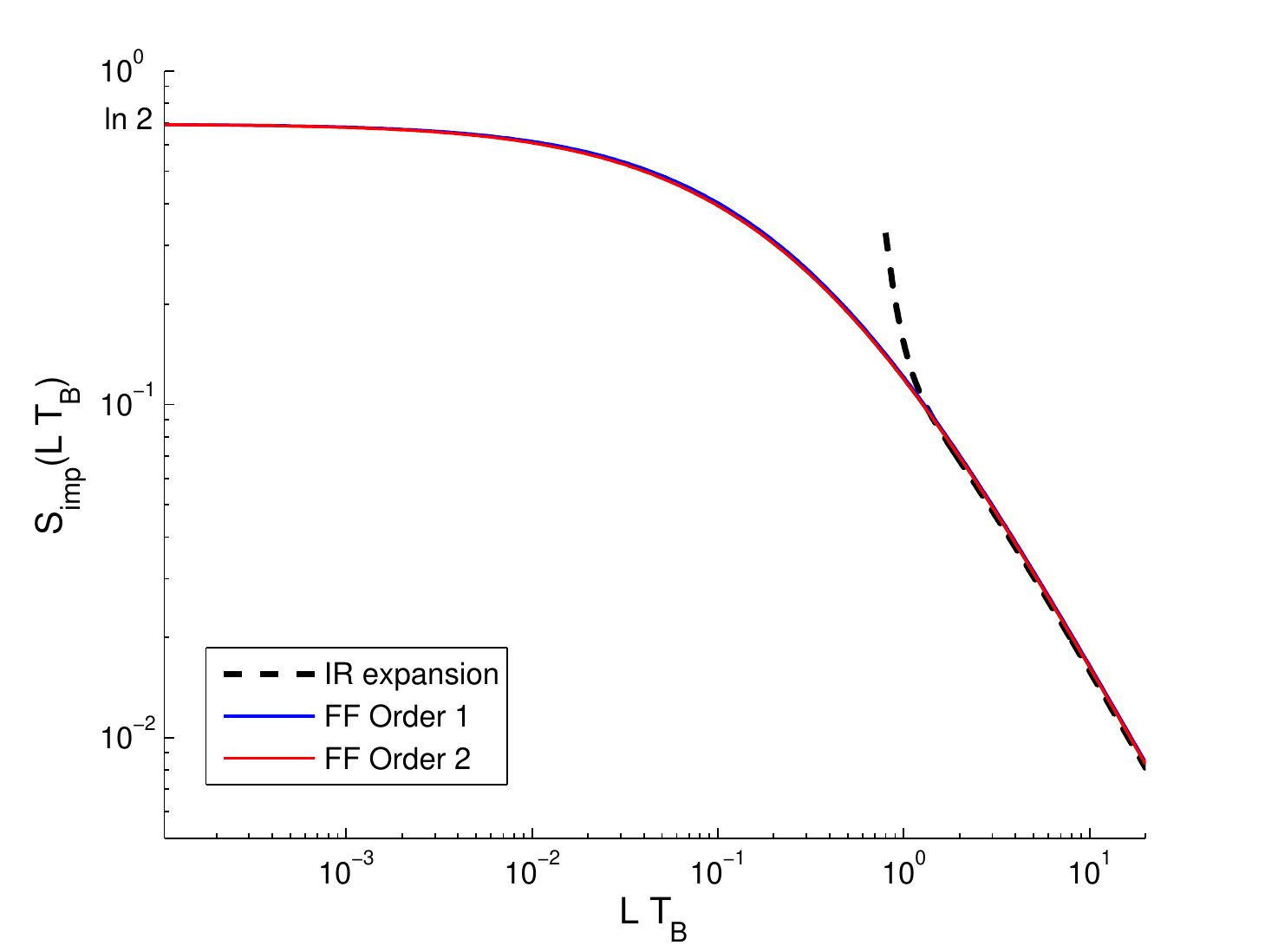}
    \caption{The form factor approximations together with the IR expansion}
\label{FFIR}
\end{figure}

\begin{figure}
\centering
    \includegraphics[scale=.45]
    {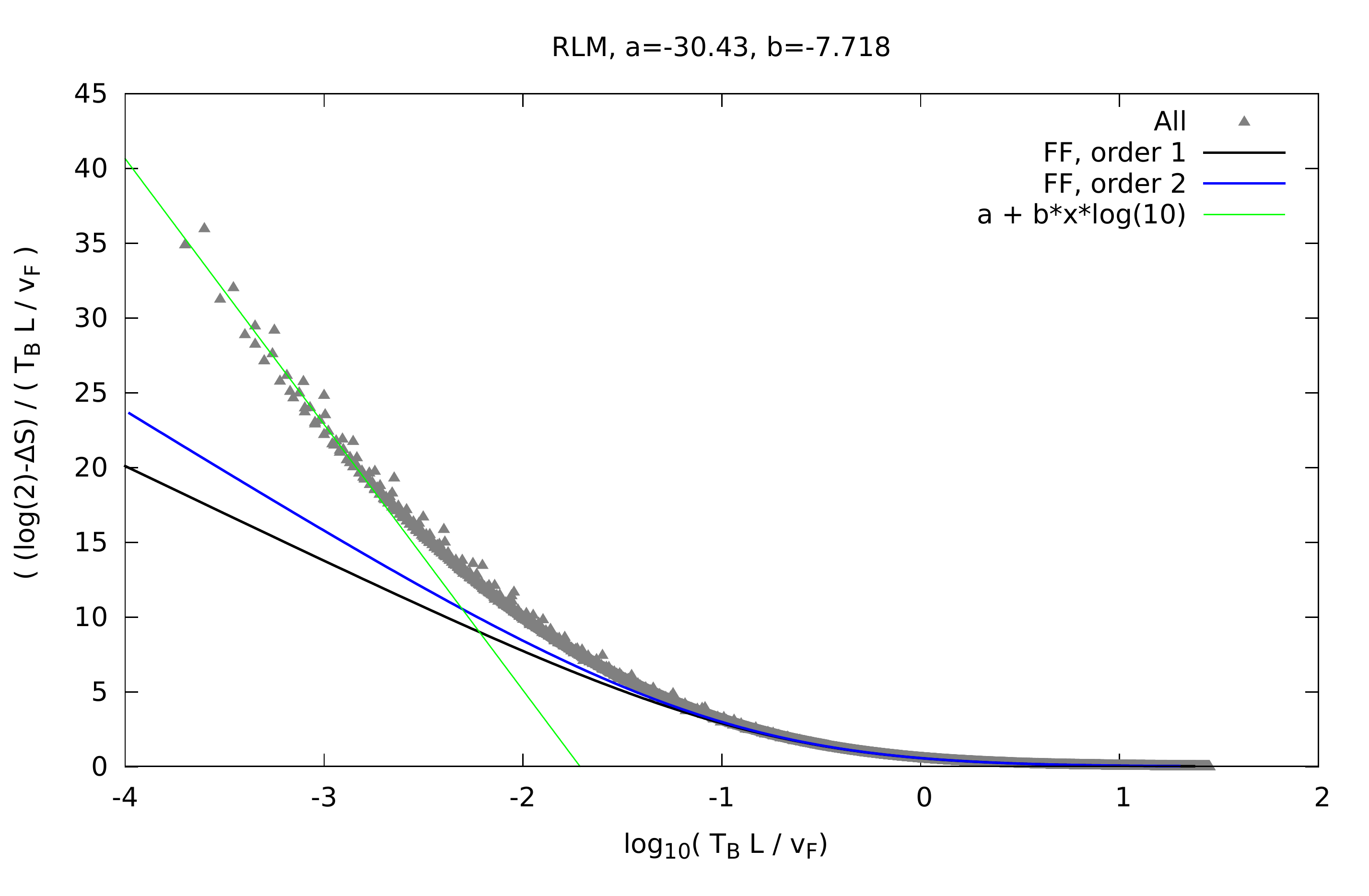}
    \caption{Singularity in the  UV, with $\Delta S \equiv  {\cal S}_{\rm imp}(LT_B) $}
\label{Peter3}
\end{figure}

We plot our results for the FF approach on figure~\ref{FFIR} where the dashed line is the IR expansion (see above and \cite{FretonIR}), and the full colored lines are form factors approximations. Clearly, on this scale, the (renormalized) FF expansion has converged very quickly. We shall soon see how close it is to the real data from numerical simulations on the XX chain.

We now discuss briefly the UV behavior. For $s_2$, standard tables give
\begin{equation}
\mathcal{S}^{(2)}_{\rm imp}(LT_B) = \ln 2 + 4\ln 2 \times (LT_B)\left[\ln 4 + \gamma +\ln(LT_B)\right]+\ldots
\end{equation}
where $\gamma \simeq 0.5772\dots$ is the Euler constant. This provides a leading order correction in the UV which reads $4\ln 2 (LT_B)\ln (LT_B)$. Note that this is compatible with what was expected from the general discussion about the non perturbative behavior in the UV: we get a term linear in $LT_B$, decorated by  logarithmic  corrections. The next order is more difficult to handle analytically, but very accurate numerics shows that it does behave similarly, only leading to  a  correction of the  slope which goes from $4\ln 2=2.77259$ to $3.53$. We plot on figure~\ref{Peter3} the leading correction divided by $L T_B$ as a function of $-\ln(LT_B)$, together with numerical data that shall be discussed in the next section. We see that the FF expansion converges very well except in the UV, where it seems to still converge, but more slowly, giving us only a rough approximation of the exact leading singularity.

Finally, the results for the entanglement for the problem of weakly coupled XXZ chains at $\Delta=-{\sqrt{2}\over 2}$ would be identical but for an overall normalization by a factor $1/2$.

\section{Numerics}\label{sec:numerics}

We now turn to a numerical  determination of  the entanglement entropy in the RLM, going back to the formulation~(\ref{eqRLMlattice}) where we will now also have to be careful with the overall finite size of the system. We write 
\begin{equation}
H=-J\sum_{m=-M'}^{-2} \left(c_{m}^\dagger c^{}_{m+1}+\hbox{{\rm h.c.}}\right)-J\sum_{m=1}^{M'-1} \left(c_{m}^\dagger c^{}_{m+1}+\hbox{{\rm h.c.}}\right)-{\tilde J}(c_{-1}^\dagger c^{}_0+c_0^\dagger c^{}_{-1}+c_0^\dagger c^{}_1+c_1^\dagger c^{}_0)
\end{equation}
%
So the left and right leads have $M'$ sites, and the impurity sits at  site 0.
We can now switch to a representation of symmetric and antisymmetric combination of the lead sites, $C (\tilde{C}) = ( c_j \pm c_{-j})/ \sqrt{2}$. Since the antisymmetric combination decouples from the rest, its contribution to the impurity entanglement will drop out. It is therefore sufficient to study a system of $M \equiv M' + 1$ sites
\begin{equation}
 H = - J \sum^{M'-1}_{m=1}  \left( C_{m}^\dagger C_{m+1}+\hbox{{\rm h.c.}}\right) 
    \;-\; {\tilde J}\sqrt{2} \left( C^\dagger_{1}  C^{}_{0}  +  C^\dagger_{0}  C^{}_{1} \right)  \,. 
\end{equation}
Here a single resonant level couples to a single chain of $M $ sites. In order to compare
with field theory we use exactly half filled system to exploit the linear regime of the cosine band.
In return we have to use an even number of $M=M'+ 1$ sites.

The scale of the resonant level with a coupling of $J' = \sqrt{2} \tilde{J}$ is
\begin{equation}
  T_{\mathrm B} / v_{\mathrm F} = \frac{J'^2}{2 \sqrt{ 1 -  J'^2} }   \,,
\end{equation}
with $v_F=2$ as we have chosen the normalization $J=1$.
Following the recipe of \cite{PeschelEisler:2009} we now calculate the reduced single particle matrix $\rho_{{\mathrm I}, L+1}$ for the last $L+1$ sites, where the first site (labeled $0$) corresponds to the impurity. 
In order to obtain the bulk result we determine the reduced density matrix $\rho_{{\mathrm B}, L}$ for the first  $L$ sites of the chain. 
The reason for taking the bulk result from the opposite end of the chain is that we cannot just study a chain of  $M'$ sites, 
as we would then have a degenerate ground state as $M'$ is odd.
The diagonalization is performed within double precision, while the trace for the entropy is performed using quadruple (128 bit) precision.

The entanglement entropy corresponding to the single particle reduced density matrix is now given by
\begin{equation}
	S = - \trace \rho \ln\rho   - \trace \left( 1- \rho\right) \ln\left( 1- \rho\right) \,,
\end{equation}
which finally leads to
\begin{equation}
	{\cal S}_{{\mathrm{imp},L}} =  S_{{\mathrm I}, L+1} - S_{{\mathrm B},L}.
\end{equation}

\begin{figure}
\centering
    \includegraphics[scale=.45]{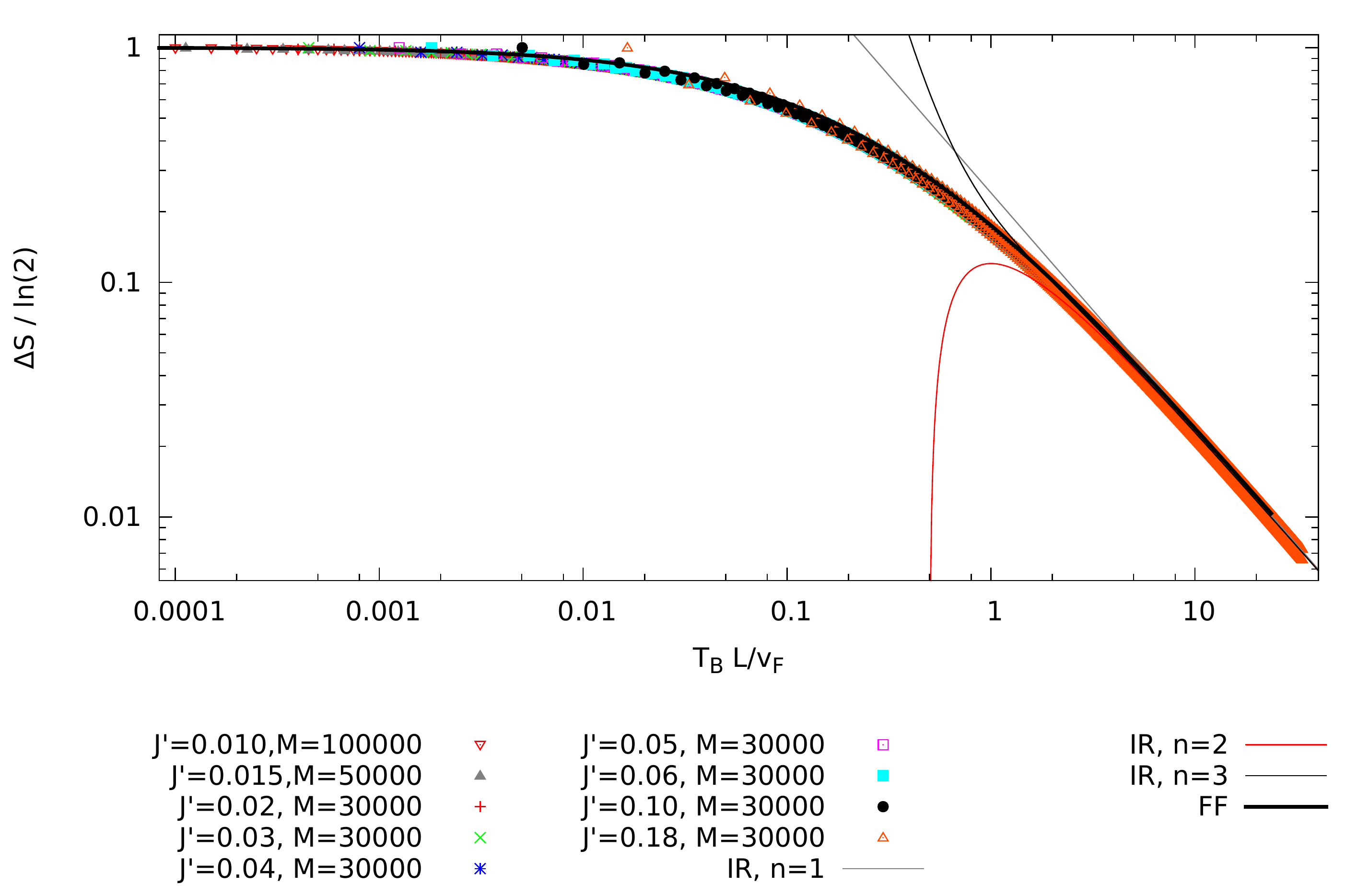}
    \caption{Comparison of numerical results and various approximations (recall that $\Delta S = {\cal S}_{\rm imp}(LT_B) $).}
\label{Peter1}
\end{figure}

In figure~\ref{Peter1} we plot the numerical results together with the first three orders of the IR expansion and the first order of the FF expansion. 
We would like to remark that on the lattice we get small $2k_{\mathrm F}$ oscillations on top of the continuum result.

\begin{figure}
\centering
    \includegraphics[scale=.45]{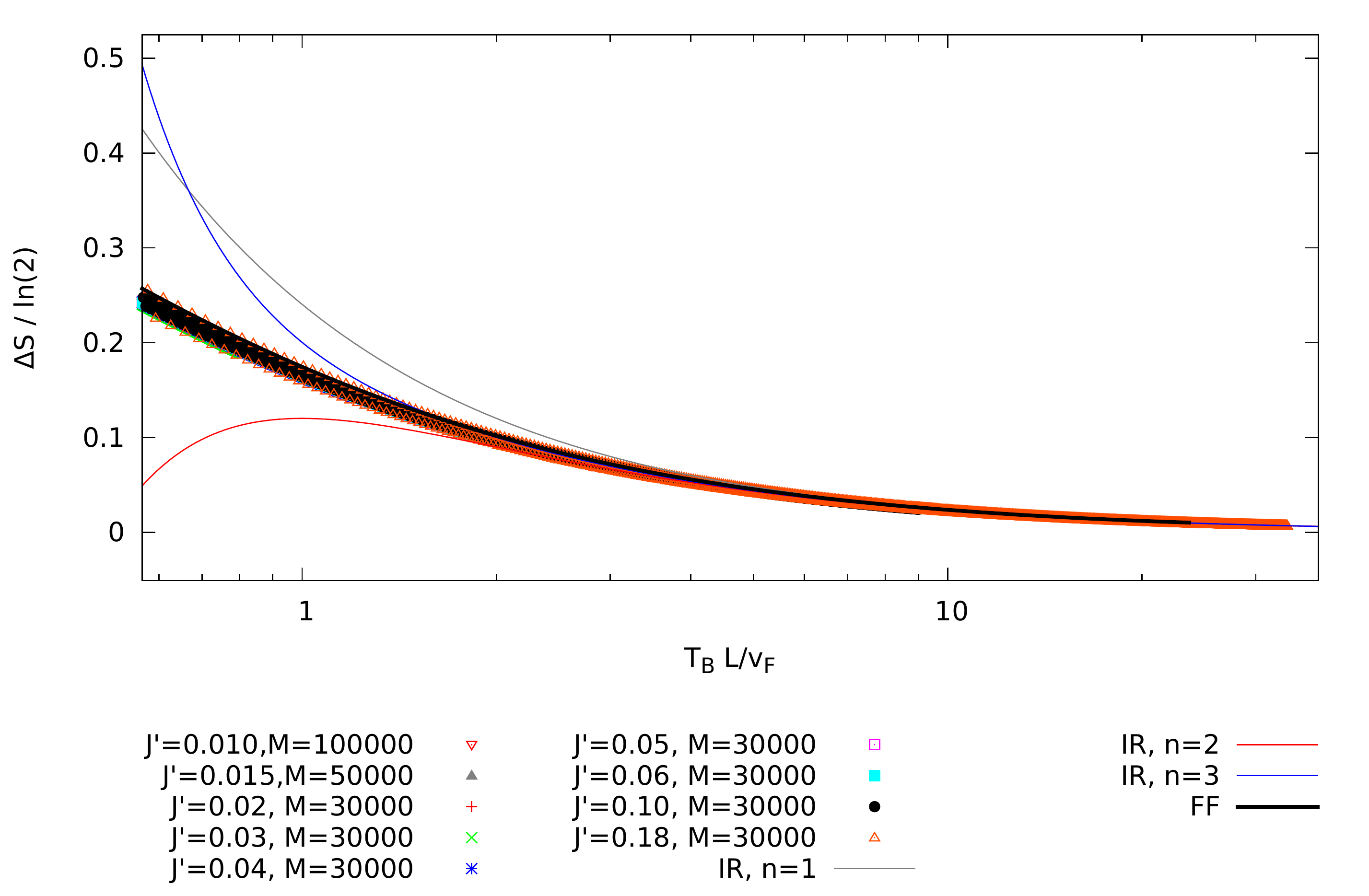}
    \caption{Comparison of numerical results and various approximations: focus on the IR}
\label{Peter2}
\end{figure}

Figure~\ref{Peter2} is a similar plot emphasizing the IR behavior.

Finally, in figure~\ref{Peter3}  -- as commented already -- we focus on the singularity in the UV, comparing slopes obtained from the FF expansion. 
Our numerics on system sizes $M=3\cdot 10^4 \ldots 10^5$ is consistent with a singularity 
\begin{equation}
{\cal S}_{\rm imp}=\ln 2 + \alpha L T_B \ln(LT_B) + \dots,
\end{equation}
with a slope $ -7.5 \lesssim\alpha \lesssim -8$. 
By applying damped boundary conditions (DBC) \cite{Bohr:PRB2007} one can access very small energy scales on system sizes which are accessible by numerics.
By looking at systems of $M=4000$ sites where we scale down the bulk hopping elements by a factor of $\Lambda=0.98$ on each bond from site 2000 to 3000 
and using a bulk hopping element of $J \Lambda^{1000}$ on the last 1000 sites we find an indication that the singularity is even slightly stronger.
While we can exclude an $L^2$ behaviour, we can not rule out the possibility of an  $L T_B \ln^2(LT_B)$ contribution.
Note that the DBC change the form of the density of states at the Fermi surface, for details see \cite{PS:AnnPh2010}.
It is therefore possible that this additional increase is due to this modification of the level spacing at the Fermi surface.
Due to the slow increase of the logarithm such a clarification is asking for multi precision arithmetic.
 
\section{Conclusion}

This study shows that the entanglement entropy of quantum impurities involved in an RG flow is a quantity which is difficult to access. It is non perturbative in the UV, and the IR perturbation, while well defined, does not capture the crossover regime very well. The form-factors approach, on the other hand, is remarkably successful. It is, however, difficult to develop except in the simple case of the Ising model, and more work will have to be done in that direction. Nevertheless, we believe that the essential features of ${\cal S}_{\rm imp}$ are under control, although it would be useful to check the UV singularities for other values of the coupling (anisotropy).

In conclusion, we emphasize that the geometry considered in this paper where the interval for the entropy is centered on the impurity is probably not the most natural physically. To characterize the Kondo physics, one would rather be interested in the entanglement of the two wires tunneling through an impurity. This could be characterized physically by the entropy of an interval with the impurity at its boundary, or for example, by the negativity of two intervals in the different wires (see {\it e.g.}~\cite{Negativity,Negativity2,Negativity3} for examples related to the Kondo problem). This situation is unfortunately much more complicated technically, mostly because the folding procedures described in this paper no longer apply. However, we still expect the conclusions of this paper to hold in that case as well, namely, we expect the entanglement entropy (or other entanglement estimators) to depend non-pertubatively on the coupling to the impurity when this is weak. We believe that improper regularizations of the IR divergences encountered in perturbation theory led to some confusion in the literature~\cite{Levine}. We will report on this -- together with a correct calculation -- in a subsequent paper.

\bigskip

\noindent {\bf Acknowledgments}:  We thank Ian Affleck, Benjamin Doyon, John Cardy and Pasquale Calabrese for discussions. HS was supported by the French Agence Nationale pour la Recherche (ANR Projet 2010 Blanc SIMI 4 : DIME) and the US Department of Energy (grant number DE-FG03-01ER45908). 
Computations were performed on the compute cluster of  the YIG 8-18 group of Peter Orth at  the Karlsruhe Institute of Technology.

\appendix

\section{Relationship between the case $h=\frac{1}{2}$ and the boundary Ising model}\label{sec:relsIsing}

Let us begin by discussing the relationship between the  RLM and  the Ising model with a boundary magnetic field. This can be seen in various ways. Start  from the hamiltonian~(\ref{RLMferm}), unfold the wires to get only right movers, and form  the combinations 
\begin{eqnarray}
\Psi_R\equiv{1\over \sqrt{2}}\left(\psi_{1R}+\psi_{2R}\right),\nonumber\\
\tilde{\Psi}_R\equiv{1\over \sqrt{2}}\left(\psi_{1R}-\psi_{2R}\right).
\end{eqnarray}
The fermion $\tilde{\Psi}_R$ decouples from the impurity entirely, and we will mostly discard it from now on. The remaining dynamics is then encoded in the hamiltonian
\begin{equation}
H=-i\int_{-\infty}^\infty \Psi_R^\dagger\partial_x\Psi_R {\rm d} x +\lambda\sqrt{2}\left[\Psi_R^\dag(0)d+\hbox{{\rm h.c.}}\right]
\end{equation}
We then refold this hamiltonian to map back to a boundary problem, introducing a $\Psi_{L}$  component:
\begin{equation}
H=-i\int_{-\infty}^0 \left[\Psi_R^\dagger\partial_x\Psi_R-\Psi_L^\dagger\partial_x\Psi_L\right]{\rm d}x+\lambda\sqrt{2}\left[\Psi^\dag(0)d+\hbox{{\rm h.c.}}\right]
\end{equation}
where $\Psi(0)\equiv \Psi_L(0)=\Psi_R(0)$. The next -- and almost final step -- is to go to a Majorana version of this problem. We decompose the fermions into real and imaginary parts as
\begin{eqnarray}
\Psi_R={1\over \sqrt{2}}(\xi_R+i\eta_R)\nonumber\\
\Psi_L={1\over\sqrt{2}}(\xi_L+i\eta_L)
\end{eqnarray}
where $\xi,\eta$ are real and obey $\{\xi_R(x),\xi_R(x')\}= \delta(x-x')$ etc. We set similarly $d={a+ib\over \sqrt{2}}$ with $\{a,a\}=\{b,b\}=1$. The problem then decouples into {\sl two independent} Majorana problems $H=H_1+H_2$, with
\begin{eqnarray}
H_1=-{i\over 2}\int_{-\infty}^0\left[ \xi_R\partial_x\xi_R-\xi_L\partial_x\xi_L\right]{\rm d}x+ \frac{i}{\sqrt{2}} \lambda \xi(0) b, \nonumber\\
H_2=-{i\over 2}\int_{-\infty}^0\left[ \eta_R\partial_x\eta_R-\eta_L\partial_x\eta_L\right]{\rm d}x- \frac{i}{\sqrt{2}} \lambda\eta(0)a, \label{Ising}
\end{eqnarray}
and $\xi(0)\equiv \xi_R(0) + \xi_L(0)$, same for $\eta$. The problems correspond of course to two Ising models with a boundary magnetic field  proportional to $\pm \lambda$ (up to normalizations), the boundary spin operator being $\sigma_B(0)= i (\xi_R+\xi_L)(0)b$. 

Note that this result is compatible with boundary entropy counting. The flow from UV to IR in the original problem leads to $g_{\rm UV}/g_{\rm IR}=2$ since a dot with two states is screened. In each of the Ising models meanwhile, we have a flow from free to fixed, with $g_{\rm free}/g_{\rm fixed}=\sqrt{2}$, so the product of the two ratios -- one for each Ising copy -- is $2$ indeed. 

Turning now to entanglement entropy, we see that the RLM entanglement for a region of size $2L$ centered around the impurity is exactly twice the entanglement for a region of length $L$ on the edge of the system in the boundary Ising model. This has been studied numerically {\it e.g.} in~\cite{Zhou,AffleckLaflorencie2}, and we will present in section~\ref{sec:FF} an analytical calculation of this quantity. 

The boundary sine-Gordon problem at $\mu={1\over 2}$ is also well known to be equivalent to two boundary Ising models~\cite{LLS}, where this time only one of these models experiences a non zero boundary magnetic field. Hence, we shall also be able to obtain the entanglement entropy for~(\ref{firstH}) for this value of $\mu$, that is $\Delta=-{\sqrt{2}\over 2}$.


%
%
%
    

\begin{thebibliography}{99}



\bibitem{Doyon1} J. Cardy, O. Castro-Alvaredo and B. Doyon, {\sl Form factors of branch-point twist fields in quantum integrable models and entanglement entropy}, J. Stat. Phys. {\bf 130}, 129--168 (2008).

\bibitem{Doyon} O. Castro-Alvaredo and B. Doyon, {\sl Bipartite entanglement entropy in massive QFT with a boundary: the Ising model}, J. Stat. Phys. {\bf 134}, 105--145 (2009).

\bibitem{AffleckLaflorencie1} E.S. S{\o}rensen, M.-S. Chang, N. Laflorencie and I. Affleck, {\sl Quantum impurity entanglement}, J. Stat. Mech. P08003 (2007).

\bibitem{AffleckLaflorencie2} I. Affleck, N. Laflorencie and  E.S. S{\o}rensen, {\sl Entanglement entropy in quantum impurity systems and systems with boundaries}, J. Phys. A: Math. Theor. {\bf 42}, 504009 (2009).
 

\bibitem{ChoKenzie} S.Y. Cho and R.H. Mc Kenzie, {\sl Quantum entanglement in the two-impurity Kondo model}, Phys. Rev. A {\bf 73}, 012109 (2006).
 
 

\bibitem{Affleck} I. Affleck, {\sl The Kondo screening cloud: what it is and how to observe it}, {\tt arXiv:0911.2209}.
  
    

\bibitem{LeHur1} A. Kopp and K. Le Hur, {\sl Universal and Measurable Entanglement Entropy in the Spin-Boson Model}, Phys. Rev. Lett. {\bf 98}, 220401 (2007).
 
 

\bibitem{LeHur2} K. Le Hur, P. Doucet-Beaupre and W. Hofstetter, {\sl Entanglement and Criticality in Quantum Impurity Systems}, Phys. Rev. Lett. {\bf 99}, 126801 (2007).
 
 

\bibitem{LeHur3} K. Le Hur, {\sl Entanglement entropy, decoherence, and quantum phase transitions of a dissipative two-level system}, Ann. of Physics {\bf 323}, 2208 (2008).
 
 
  

\bibitem{CardyCalabrese1} P. Calabrese and J. Cardy, {\sl Entanglement Entropy and Quantum Field Theory}, J. Stat. Mech. 0406:P06002, (2004).

\bibitem{CardyCalabrese2} P. Calabrese and J. Cardy, {\sl Entanglement entropy and conformal field theory}, J. Phys. A: Math. Gen. {\bf 42}, 504005 (2009).
  

\bibitem{AffleckBarzykin} V. Barzykin and I. Affleck, {\sl Impurity correlations in dilute Kondo alloys}, Phys. Rev. B {\bf 61}, 6170 (2000).

\bibitem{FretonIR} L. Freton, E. Boulat and H. Saleur, {\sl Infra-red expansion of entanglement entropy in the Interacting Resonant Level Model}, {\tt arXiv:1301.6535}.

\bibitem{AffleckLudwig} I. Affleck and A.W.W. Ludwig, {\sl Universal noninteger ``ground-state degeneracy'' in critical quantum systems}, Phys. Rev. Lett. {\bf 67}, 161--164 (1991).


\bibitem{AffleckEggert}  S. Eggert and I. Affleck, {\sl Magnetic impurities in half-integer-spin Heisenberg antiferromagnetic chains}, Phys. Rev. B {\bf 46}, 10866 (1992).

\bibitem{WenFQHE} X.G. Wen, {\sl Chiral Luttinger liquid and the edge excitations in the fractional quantum Hall states}, Phys. Rev. B {\bf 41}, 12838 (1990); {\bf 43}, 11025 (1991); {\bf 44}, 5708 (1991).

\bibitem{ConductanceFQHE} P. Fendley, A.W.W. Ludwig and H. Saleur, {\sl Exact Conductance through Point Contacts in the $\nu = 1/3$ Fractional Quantum Hall Effect}, Phys. Rev. Lett. {\bf 74}, 3005--3008 (1995).

\bibitem{Holzhey} C. Holzhey, F. Larsen and F. Wilczek, {\sl Geometric and renormalized entropy in conformal field theory}, Nucl. Phys. B {\bf 424}, 443--467 (1994).

\bibitem{KL} R. Konik and A. Leclair, {\sl Short distance expansions of correlation functions in the sine-Gordon theory}, Nucl. Phys. B {\bf 479} 619--653 (1996).

\bibitem{CZ} R. Chatterjee and A. Zamolodchikov, {\sl Local magnetization in critical Ising model with boundary magnetic field}, Mod. Phys. Lett. A {\bf 9}, 2227--2234 (1994).
  
  
    

\bibitem{LLS} A. Leclair, F. Lesage and H. Saleur, {\sl Exact Friedel oscillations in the $g=1/2$ Luttinger liquid}, Phys. Rev. B {\bf 54}, 13597 (1996).


\bibitem{Zamo} A. Zamolodchikov, {\sl Two-point correlation functions in scaling Lee-Yang model}, Nucl. Phys. B {\bf 348}, 619--641 (1991).

\bibitem{FLZZ} V. Fateev, S. Lukyanov, A. Zamolodchikov and Al. B. Zamolodchikov, {\sl Expectation values of boundary fields in the boundary sine-Gordon model}, Phys. Lett. B {\bf 406} 83--88 (1997).
    

\bibitem{SM} E. Sela and A. K. Mitchell, {\sl Local magnetization in the boundary Ising model at finite temperature},  J. Stat. Mech. P04006 (2012).
    
    

\bibitem{Egger} R. Egger and H. Grabert, {\sl Friedel Oscillations for Interacting Fermions in One Dimension}, Phys. Rev. Lett. {\bf 75}, 3505 (1995), {\sl Friedel Oscillations in Luttinger Liquids}, Quantum Transport in Semiconductor Submicron Structures NATO ASI Series {\bf 326}, 133--158 (1996). 

\bibitem{CCT} P. Calabrese, J. Cardy and E. Tonni, {\sl Entanglement entropy of two disjoint intervals in conformal field theory: II }, J. Stat. Mech. P01021 (2011).
    

\bibitem{ImpEnt} E.S. S{\o}rensen, M.-S. Chang, N. Laflorencie and I. Affleck, {\sl Impurity entanglement entropy and the Kondo screening cloud}, J. Stat. Mech. L01001 (2007).

\bibitem{Eriksson} E. Eriksson and H. Johannesson, {\sl Impurity entanglement entropy in Kondo systems from conformal field theory}, Phys. Rev. B {\bf 84}, 041107(R) (2011). 

\bibitem{Zhou} H.-Q. Zhou, T. Barthel, J. Fjaerestad and U. Schollw\"ock, {\sl Entanglement and boundary critical phenomena}, Phys. Rev. A {\bf 74}, 050305 (2006). 
 

\bibitem{Skorik} F. Lesage,  H. Saleur and S. Skorik, {\sl Form factors approach to current correlations in one dimensional systems with impurities}, Nucl. Phys. B {\bf 474}, 602--640 (1996).
    

\bibitem{LS}  F. Lesage and H. Saleur, {\sl Form factor computation of Friedel oscillations in Luttinger liquids}, J. Phys. A: Math. Gen. {\bf 30} L457 (1997).

\bibitem{GZ} S. Ghoshal and A.B. Zamolodchikov, {\sl  Boundary S-Matrix and Boundary State in Two-Dimensional Integrable Quantum Field Theory}, Int. J. Mod. Phys. A {\bf 9}, 3841--3886 (1994).

\bibitem{PeschelEisler:2009}  I. Peschel and V. Eisler, {\sl Reduced density matrices and entanglement entropy in free lattice models}, J. Phys. A: Math. Theor. {\bf 42}, 504003 (2009).

\bibitem{Bohr:PRB2007} D. Bohr and P. Schmitteckert, {\sl Strong enhancement of transport by interaction on contact links}, Phys.~Rev.~B {\bf 75}, 241103(R) (2007).

\bibitem{PS:AnnPh2010} A.~Bransch{\"a}del, G.~Schneider, and P.~Schmitteckert, {\sl Conductance of inhomogeneous systems: Real-time dynamics}, Annalen der Physik {\bf 522}, 657 (2010).


\bibitem{Negativity} A.~Bayat, P.~Sodano, and S.~Bose, {\sl Negativity as the entanglement measure to probe the Kondo regime in the spin-chain Kondo model}, Phys. Rev. B {\bf 81}, 064429 (2010).

\bibitem{Negativity2} A.~Deschner and E.S.~S{\o}rensen, {\sl Impurity entanglement in the $J$-$J_2$-$\delta$ quantum spin chain}, J. Stat. Mech. P10023 (2011).

\bibitem{Negativity3} A.~Bayat, S.~Bose, P.~Sodano, and H. Johannesson, {\sl Entanglement Probe of Two-Impurity Kondo Physics in a Spin Chain}, Phys. Rev. Lett. {\bf 109}, 066403 (2012).


\bibitem{Levine} G.C. Levine, {\sl Entanglement Entropy in a Boundary Impurity Model}, Phys. Rev. Lett. {\bf 93}, 266402 (2004). 


\end{thebibliography}
    \end{document}